\documentclass[a4paper,11pt]{article}
\usepackage{pos}
\usepackage{slashed}

\newcommand{\lb}{\left(}
\newcommand{\rb}{\right)}
\newcommand{\be}{\beta}
\newcommand{\al}{\alpha}
\newcommand{\lam}{\lambda}
\newcommand{\GeV}{{\ensuremath\rm GeV}}
\newcommand{\TeV}{{\ensuremath\rm TeV}}
\newcommand{\fb}{{\ensuremath\rm fb}}

\title{BSM scenarios with missing energy
at future lepton colliders}

\author*[a]{Tania Robens}

\affiliation[a]{Rudjer Boskovic Institute,\\
  Bijenicka cesta 54, 10000 Zagreb, Croatia}

\emailAdd{trobens@irb.hr}

\abstract{I will briefly discuss the signatures and discovery prospects of several new physics models containing dark matter candidates at future lepton colliders. In particular, I will discuss the two models that, among other signatures, lead to electroweak gauge bosons and missing energy: the Inert Doublet Model, as well as the THDMa, a two Higgs doublet model with an additional pseudoscalar that serves as a portal to the dark sector. Results are mainly based on a Snowmass Whitepaper \cite{Kalinowski:2022fot}.\\
RBI-ThPhys-2022-42}

\FullConference{%
  41st International Conference on High Energy physics - ICHEP2022\\
  6-13 July, 2022\\
  Bologna, Italy
}

\begin{document}
\maketitle

\section{Introduction}
I briefly discuss two models that enhance the scalar sector of the Standard Model by additional particle content, including dark matter candidates. These models lead to signatures with missing energy. I present current bounds on these models as well as perspectives and rates at future lepton colliders.
\section{Inert Doublet Model}

The Inert Doublet Model (IDM) \cite{Deshpande:1977rw,Cao:2007rm,Barbieri:2006dq} is a two Higgs Doublet Model (THDM) with a discrete exact $\mathbb{Z}_2$ symmetry containing a dark matter candidate.
The model features 7 free parameters, which we chose in the so-called physical basis {\cite{Ilnicka:2015jba}}:
 $v,M_h,M_H, M_A, M_{H^{\pm}}, \lam_2, \lam_{345}$,
where the $\lam$s correspond to potential parameters. As two parameters (the vacuum expecation value (vev) $v$ and $M_h\,\sim\,125\,\GeV$) are fixed by experimental measurements, we end up with a total number of 5 free parameters. Here, we consider the case where $H$ is the dark matter candidate, which implies $M_{A,\,H^\pm}\,\geq\,M_H$.

\begin{figure}[htb]
\centerline{%
\includegraphics[width=0.48\textwidth]{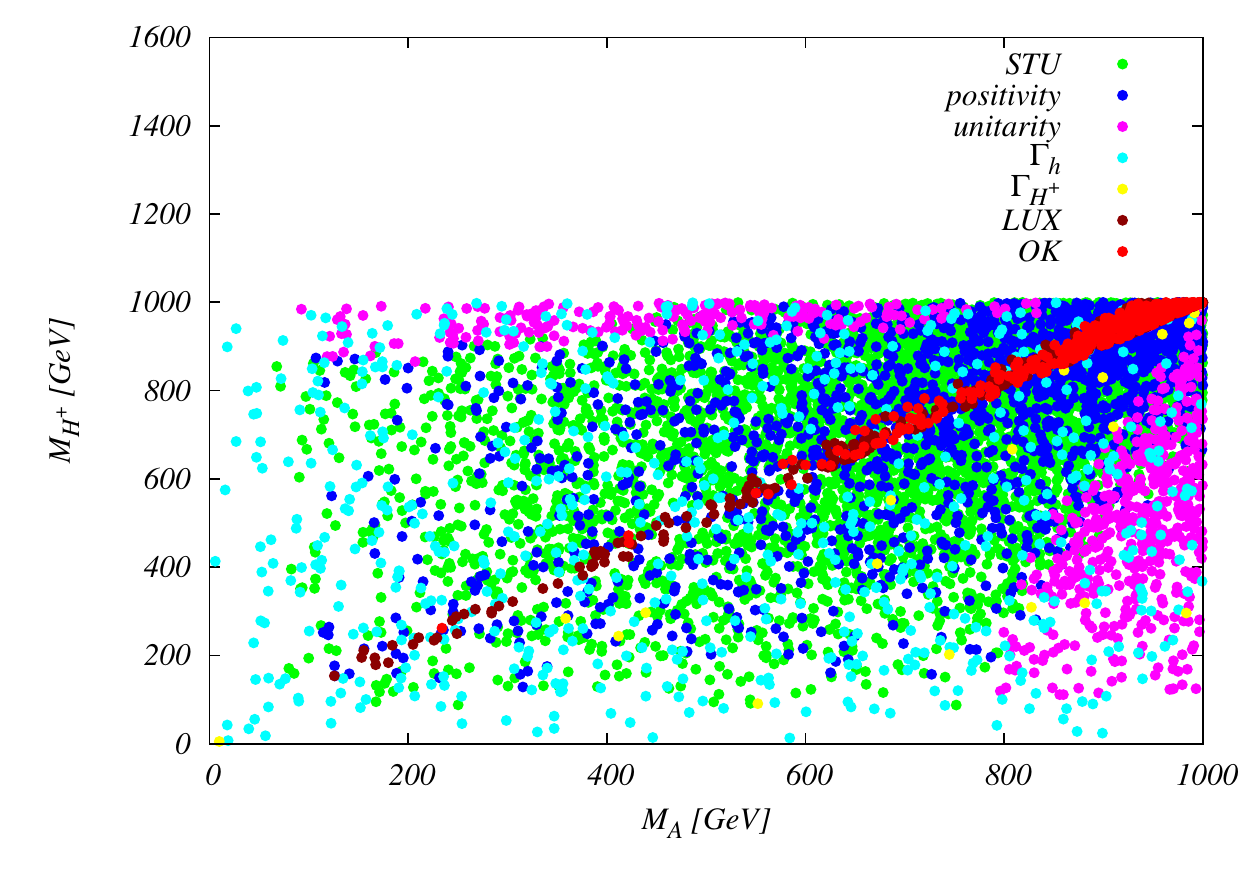}
\includegraphics[width=0.48\textwidth]{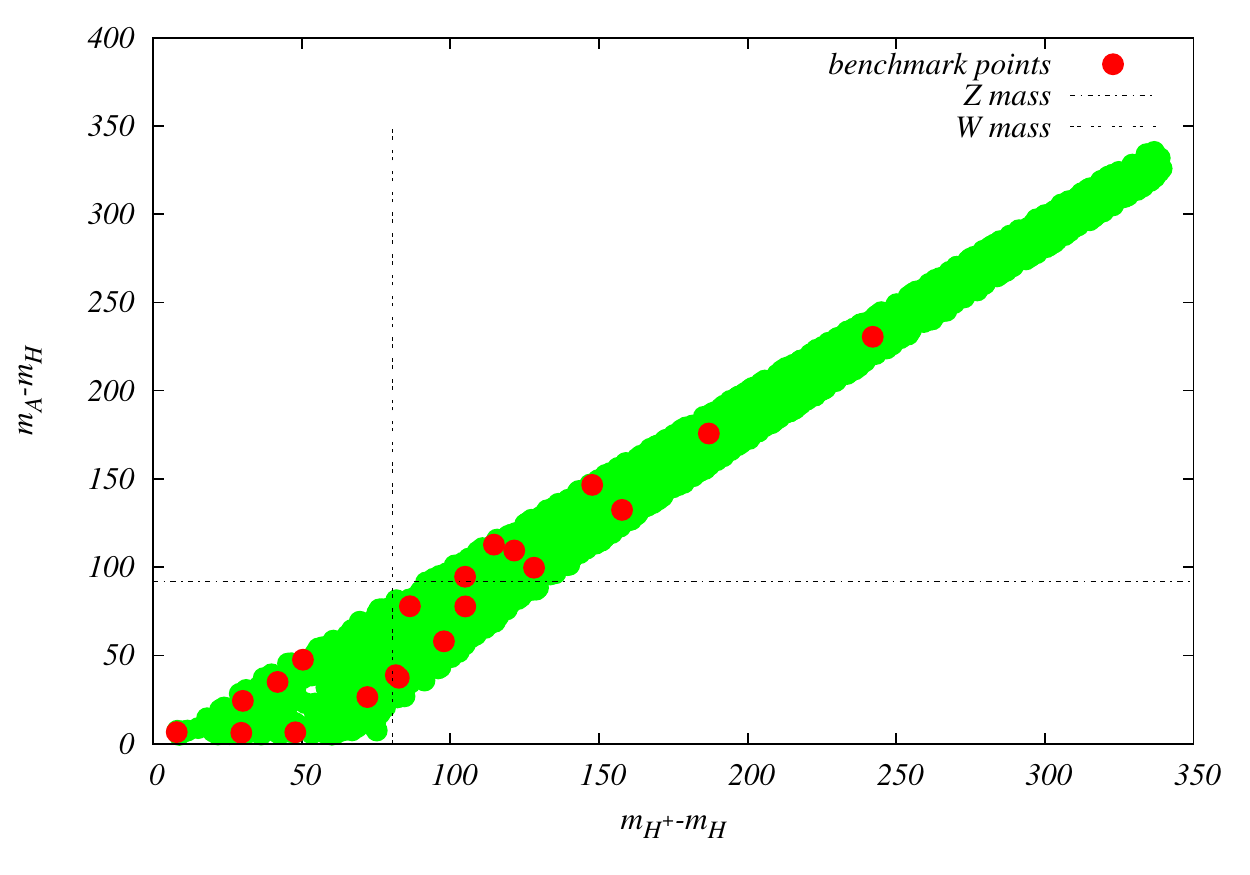}}
\caption{Masses are requested to be quite degenerate after all constraints have been taken into account. {\sl Left:} In the $\lb M_A,\,M_{H^\pm} \rb$ plane (taken from \cite{Ilnicka:2015jba}). {\sl Right:} In the {$\lb M_{H^\pm}-M_H,\,M_A-M_H \rb$} plane (taken from \cite{Kalinowski:2018ylg}).}
\label{fig:massesidm}
\end{figure}

\begin{figure}[htb]
\centerline{%
\includegraphics[width=0.48\textwidth]{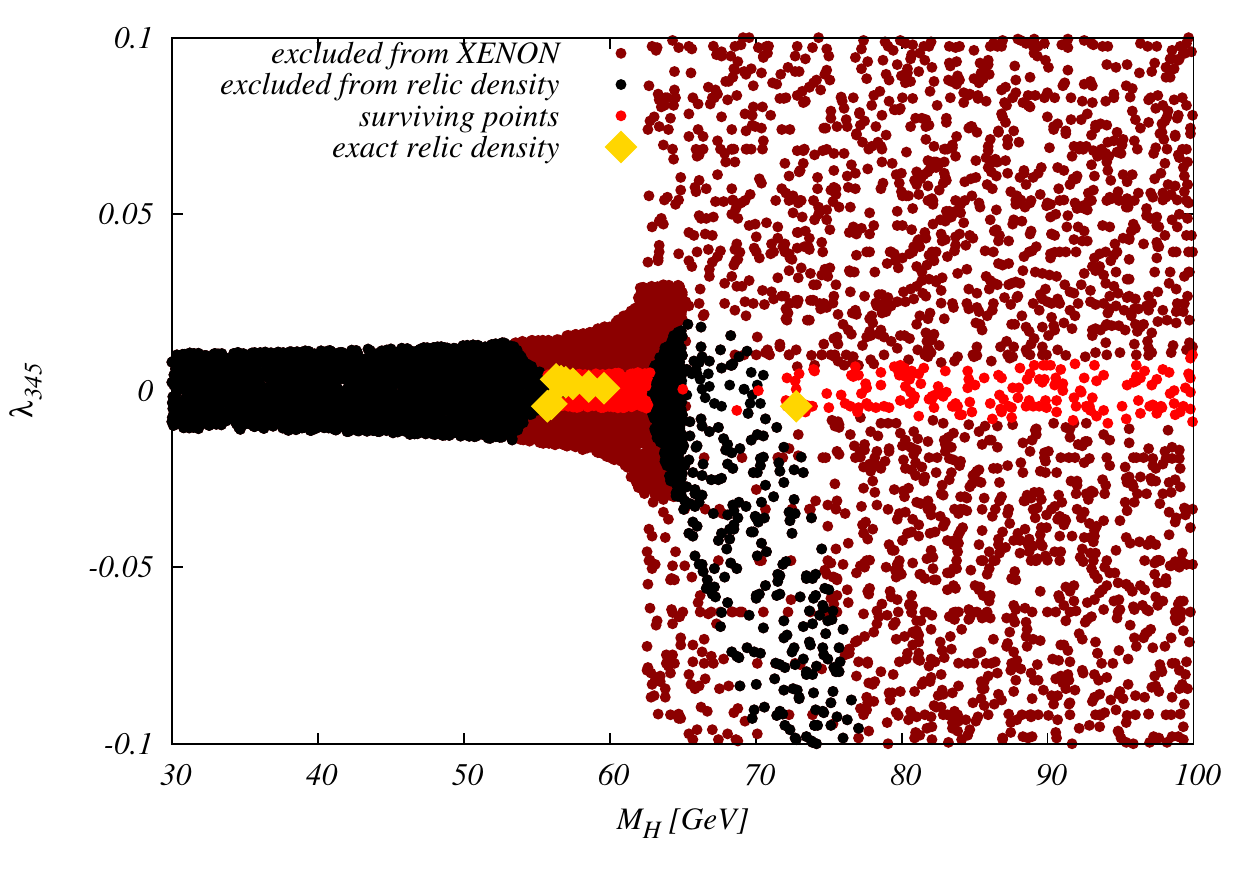}}
\caption{Interplay of signal strength and relic density constraints in the $\lb M_H,\,\lam_{345}\rb$ plane. Using XENON1T results, with golden points labelling those points that produce exact relic density (taken from \cite{Ilnicka:2018def}). Note that all points displayed here also pass the new LUX-Zeppelin bounds \cite{LZ:2022ufs}.}
\label{fig:lowmh}
\end{figure}

The model is subject to a large number of theoretical and experimental constraints \cite{Ilnicka:2015jba,Ilnicka:2018def,Dercks:2018wch,Kalinowski:2018ylg,Kalinowski:2020rmb,Kalinowski:2022fot}. 
These lead to a large reduction of the allowed parameter space. As an example, the masses are usually quite degenerate, as can be seen {from} figure \ref{fig:massesidm}\footnote{Note that BP11 from \cite{Kalinowski:2018ylg} is by now excluded from the newest direct detection constraints \cite{LZ:2022ufs}.}. 
We also consider the case when $M_H\,\leq\,M_h/2$, where constraints from  $h\,\rightarrow\,\text{inv{isible}}$ start to play an important role and an interesting interplay arises, between bounds from signal strength measurements, 
and bounds from dark matter relic density, 
see figure \ref{fig:lowmh}. In \cite{Ilnicka:2015jba}, it was found that this in general leads to a lower bound of $M_H\,\sim\,50\,\GeV$, with exceptions presented in \cite{Kalinowski:2020rmb}. 
The discovery potential of ILC and CLIC was investigated in \cite{Kalinowski:2018kdn,CLIC:2018fvx,Zarnecki:2019poj,Zarnecki:2020swm,Sokolowska:2019xhe,Klamka:2022ukx} for several benchmark points proposed in \cite{Kalinowski:2018ylg}, for varying center-of-mass energies from $250\,\GeV$ up to $3\,\TeV$. We focus on $AH$ and $H^+ H^-$ production with $A\,\rightarrow\,Z\,H$ and $H^\pm\,\rightarrow\,W^\pm H$, with leptonic decays of the electroweak gauge bosons. Event generation was done using \texttt{WHizard 2.2.8} \cite{Moretti:2001zz,Kilian:2007gr}, with an interface via \texttt{SARAH} \cite{Staub:2015kfa} and \texttt{SPheno 4.0.3} \cite{Porod:2003um,Porod:2011nf} for model implementation. For CLIC results  energy spectra \cite{Linssen:2012hp} were also taken into account.

\begin{figure}[htb]
\begin{center}
\includegraphics[width=0.45\textwidth]{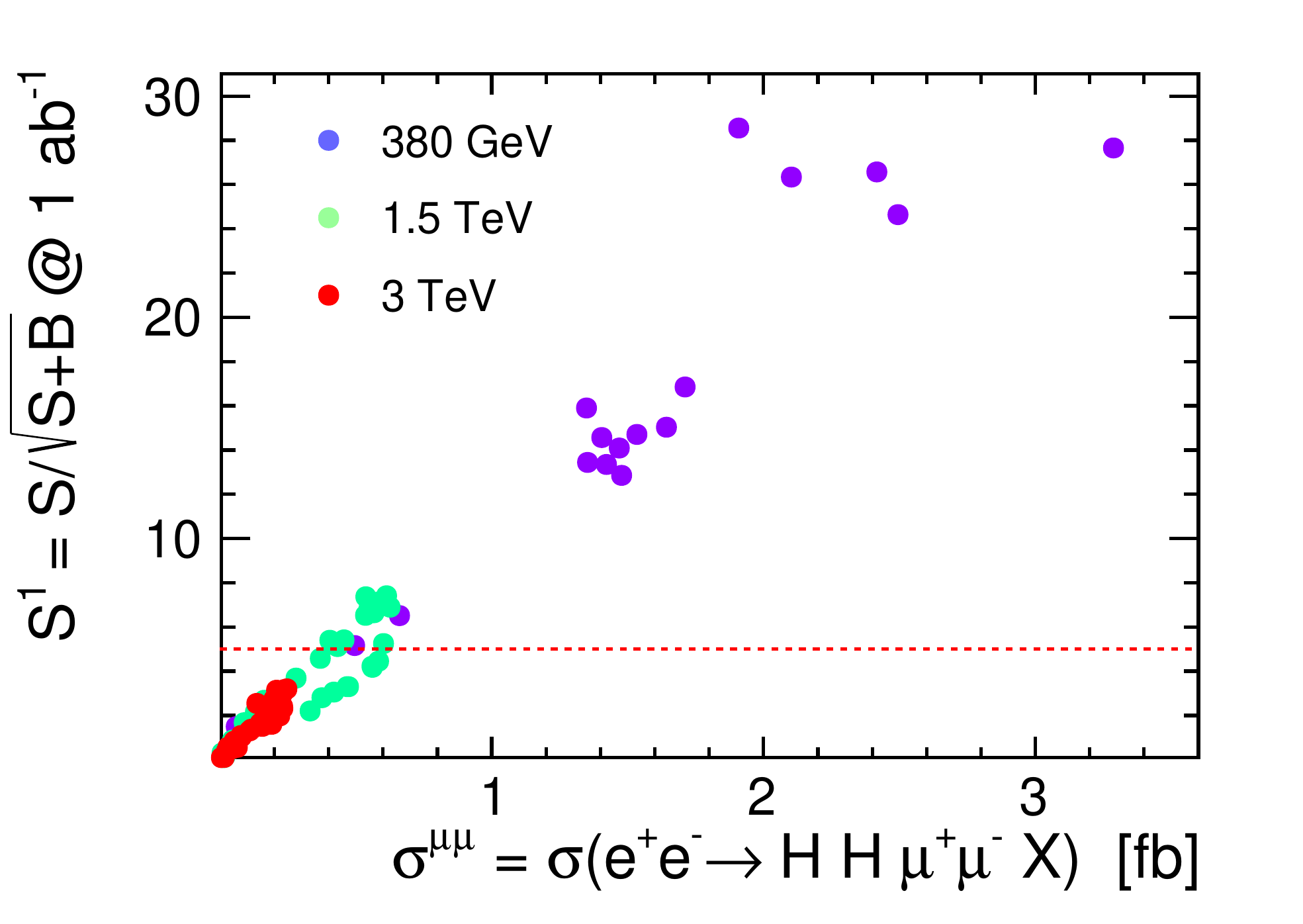}
\includegraphics[width=0.45\textwidth]{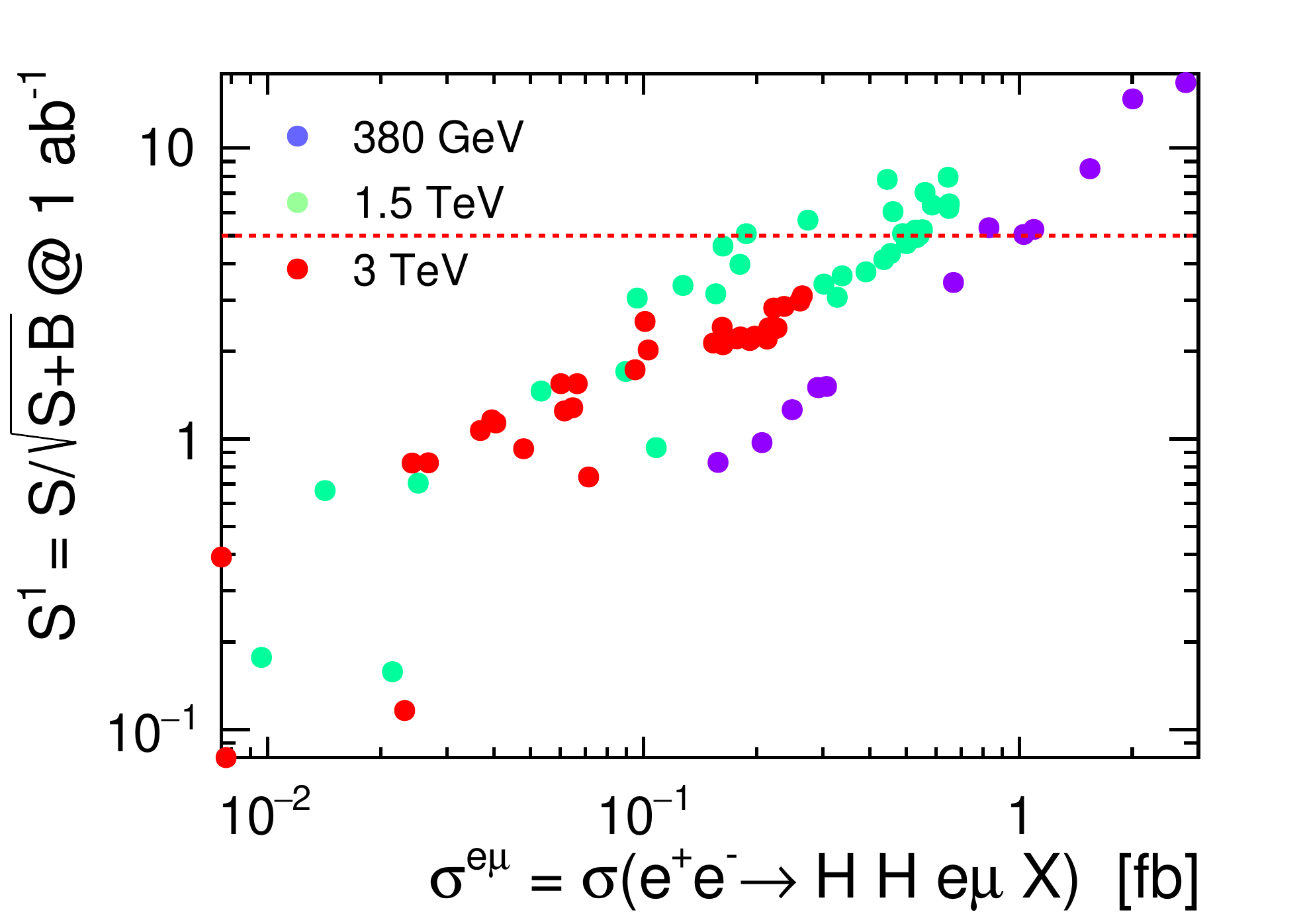}
\includegraphics[width=0.45\textwidth]{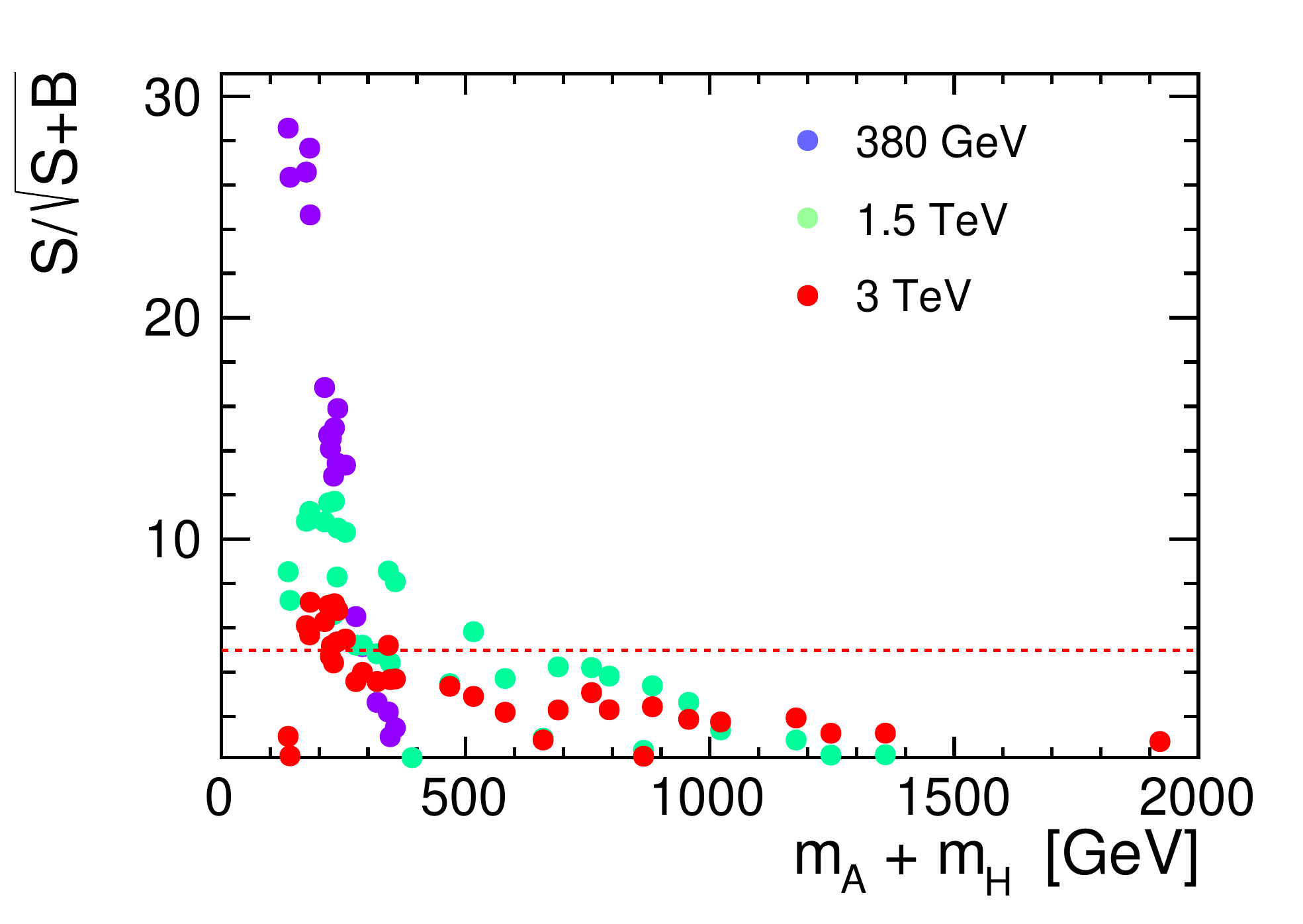}
\includegraphics[width=0.45\textwidth]{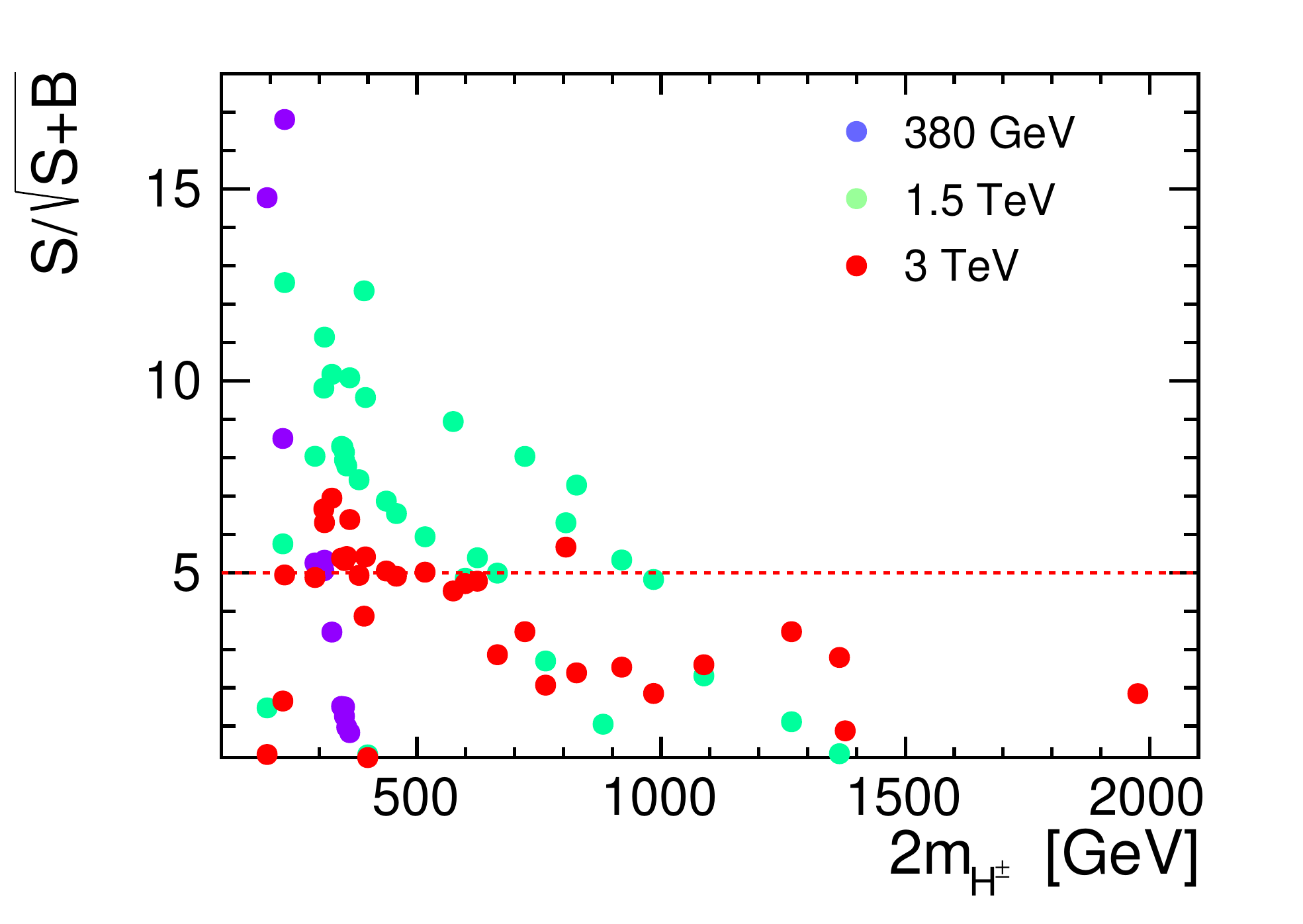}
\end{center}
\caption{Discovery prospects at CLIC for the IDM in $\mu^+\mu^-+\slashed{E}$ {\sl (left)} and $\mu^\pm\,e^\mp+\slashed{E}$ {\sl (right)} final states, as a function of the respective production cross-sections {\sl (top)} and mass sum of the produced particles {\sl (bottom)}. Taken from \cite{Kalinowski:2018kdn}.}
\label{fig:clic}
\end{figure}

The investigated final states were
$e^+\,e^-\,\rightarrow\,\mu^+\mu^-+\slashed{E},\;\;\;
e^+\,e^-\,\rightarrow\,\mu^\pm\,e^\mp+\slashed{E}$
for $HA$ and $H^+\,H^-$ production, respectively.
Results for the discovery reach of CLIC, including center-of-mass energies of 1.5\,TeV and 3\,TeV, are shown in figure \ref{fig:clic}. In general, production cross sections $\gtrsim\,0.5\,\fb$ and
mass sums up to 1 \TeV~ seem accessible, where the $\mu^\pm\,e^\mp$ channel seems to provide a larger discovery range.

\section{THDMa}

The THDMa is a type II two-Higgs-doublet model that is extended by an additional pseudoscalar $a$. In the gauge-eigenbasis, the additional scalar serves as a portal to the dark sector, with a fermionic dark matter candidate, denoted by $\chi$. More details can e.g. be found in \cite{Ipek:2014gua,No:2015xqa,Goncalves:2016iyg,Bauer:2017ota,Tunney:2017yfp,LHCDarkMatterWorkingGroup:2018ufk,Robens:2021lov}. 

The model contains the following particles in the scalar and dark matter sector: ${h,\,H,\,H^\pm}$, ${a,}\,{A,}\,{{\chi}}$. It depends on 12 additional new physics parameters

\centerline{${v,\,m_h,\,m_H,}\,{ m_a,}\,{m_A,\,m_{H^\pm},}\,{m_\chi};\;{\cos\lb \be-\al\rb,\,\tan\be,}\,{\sin\theta;\;y_\chi,}\,{\lam_3,}\,{\lam_{P_1},\,\lam_{P_2}},$}
\noindent
where $v$ and either $m_h$ or $m_H$ are fixed by current measurements in the electroweak sector. 

\begin{center}
\begin{figure}
\begin{center}
\begin{minipage}{0.45\textwidth}
\begin{center}
\includegraphics[width=\textwidth]{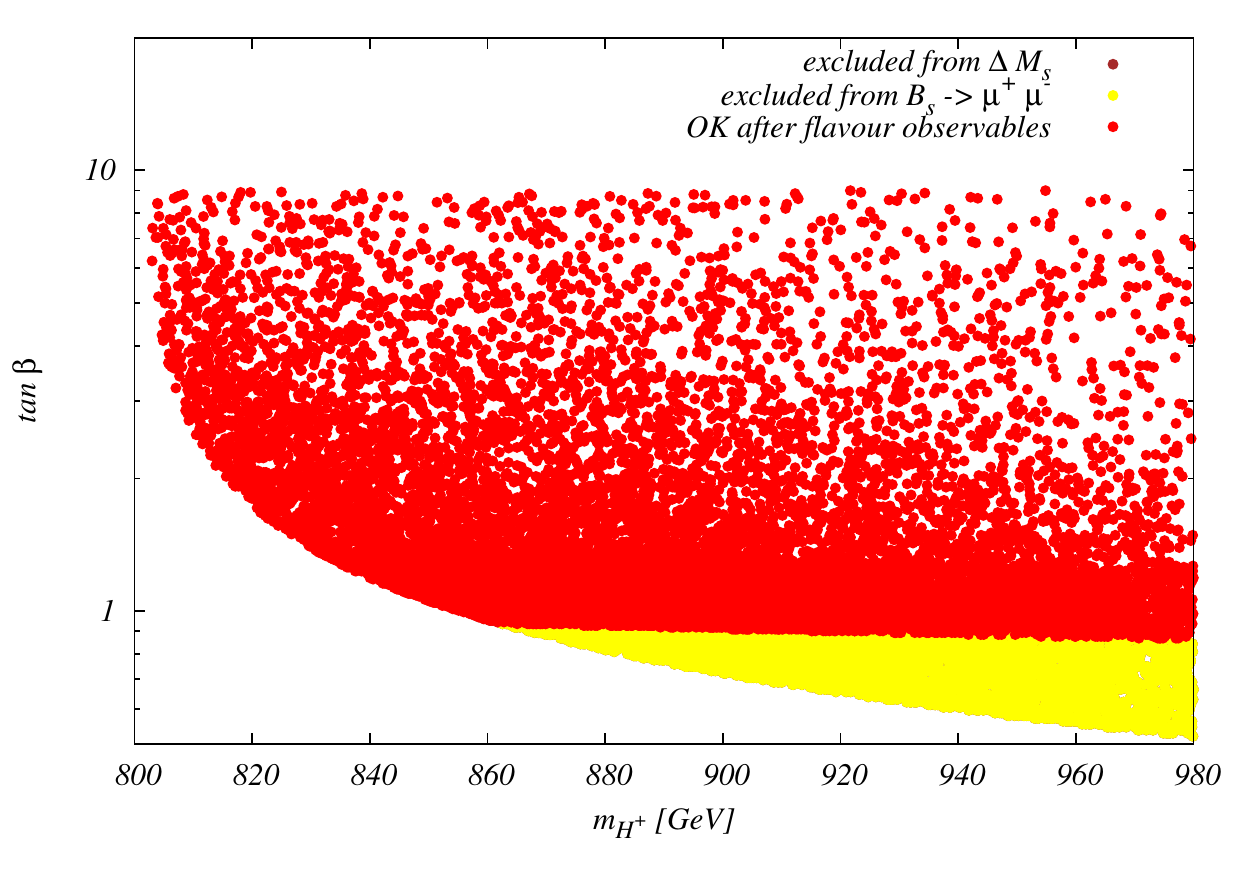}
\end{center}
\end{minipage}
\begin{minipage}{0.45\textwidth}
\begin{center}
\includegraphics[width=\textwidth]{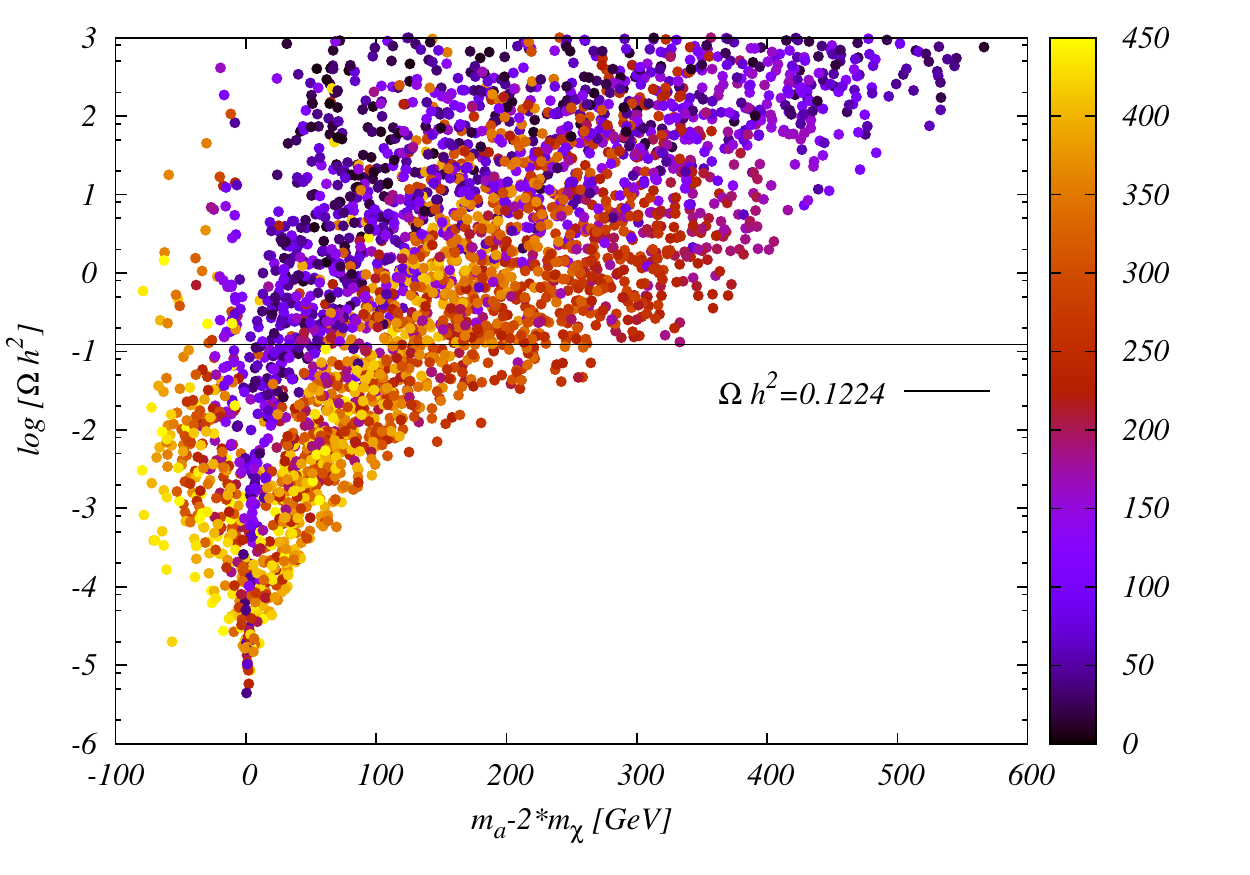}
\end{center}
\end{minipage}
\end{center}
\caption{\label{fig:thdmab} {\sl Left:} Bounds on the $\lb m_{H^\pm},\,\tan\be\rb$ plane from B-physics observables, implemented via the SPheno \cite{Porod:2011nf}/ Sarah \cite{Staub:2013tta} interface,  and compared to experimental bounds \cite{combi,Amhis:2019ckw}. The contour for low $\lb m_{H^\pm,\,\tan\be}\rb$ values stems from \cite{Misiak:2020vlo,mm}.  {\sl Right:} Dark matter relic density as a function of $m_a-2\,m_\chi$, with $m_\chi$ defining the color coding. The typical resonance-enhanced relic density annihilation is clearly visible. Right figure taken from \cite{Robens:2021lov}.}
\end{figure}
\end{center}

I here report on results of a scan that allows all of the above novel parameters float in specific predefined ranges \cite{Robens:2021lov}. 
Two examples for direct bounds in 2-dimensional planes are displayed in figure \ref{fig:thdmab}. Note that for this proceeding, on contrast to the results presented in \cite{Robens:2021lov,Kalinowski:2022fot}, we have now updated the value of $B_s\,\rightarrow\,\mu^+\,\mu^-$ to the current PDG value \cite{ParticleDataGroup:2022pth}, we have
$ B_s\,\rightarrow\,\mu^+\mu^-\,=\,\lb 3.01\,\pm\,0.35 \rb\,\times\,10^{-9}.$
Following the logic explained in \cite{Robens:2021lov}, this leads to
$ \lb B_s\,\rightarrow\,\mu^+\mu^-\rb^{\text Spheno}\,\in\,\left[1.52;3.34 \right]\,\times\,10^{-9}.$
Note that the $\Delta M_s$ experimental value has also been updated \cite{HFLAV:2022pwe} and now reads 
$\Delta M_s\lb\text{ps}^{-1}\rb\,=\,17.765\,\pm\,0.004\,\pm\,0.004.$
However, this basically leads to similar bounds as the previous value \cite{HFLAV:2019otj}, so we did not update the respective bounds.

If, for consistency, now taking again a $3\sigma$ allowed range for $B_s\,\rightarrow\,\mu^+\mu^-$, the bounds from this branching ratio and $\Delta M_s$ basically overlapp. In turn, it means that now $\tan\be$ values $\lesssim\,1$ are still allowed. 
The second plot displays the relic density as a function of the mass difference $m_a-2\,m_\chi$. 

\begin{center}
\begin{figure}
\begin{center}
\begin{minipage}{0.45\textwidth}
\begin{center}
\includegraphics[width=\textwidth]{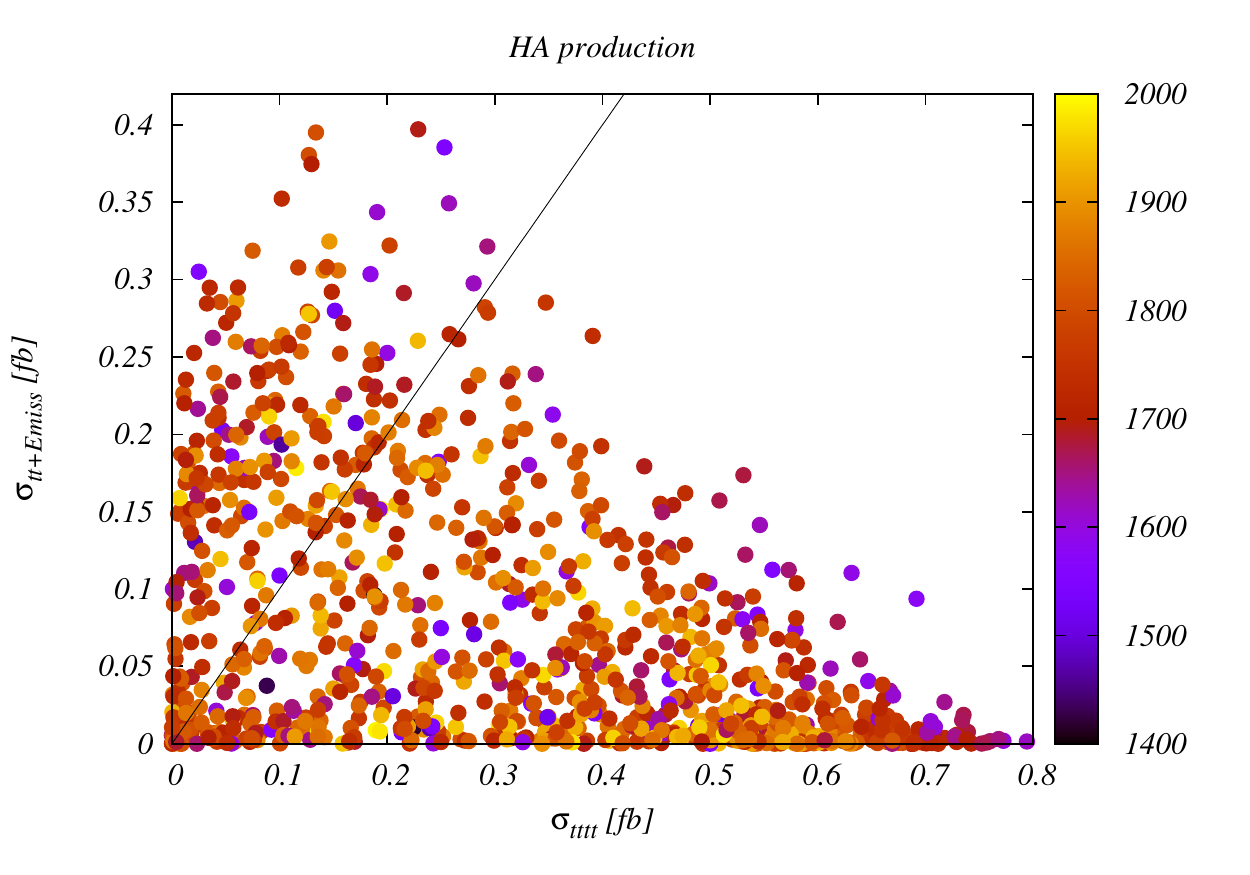}
\end{center}
\end{minipage}
\begin{minipage}{0.45\textwidth}
\begin{center}
\includegraphics[width=\textwidth]{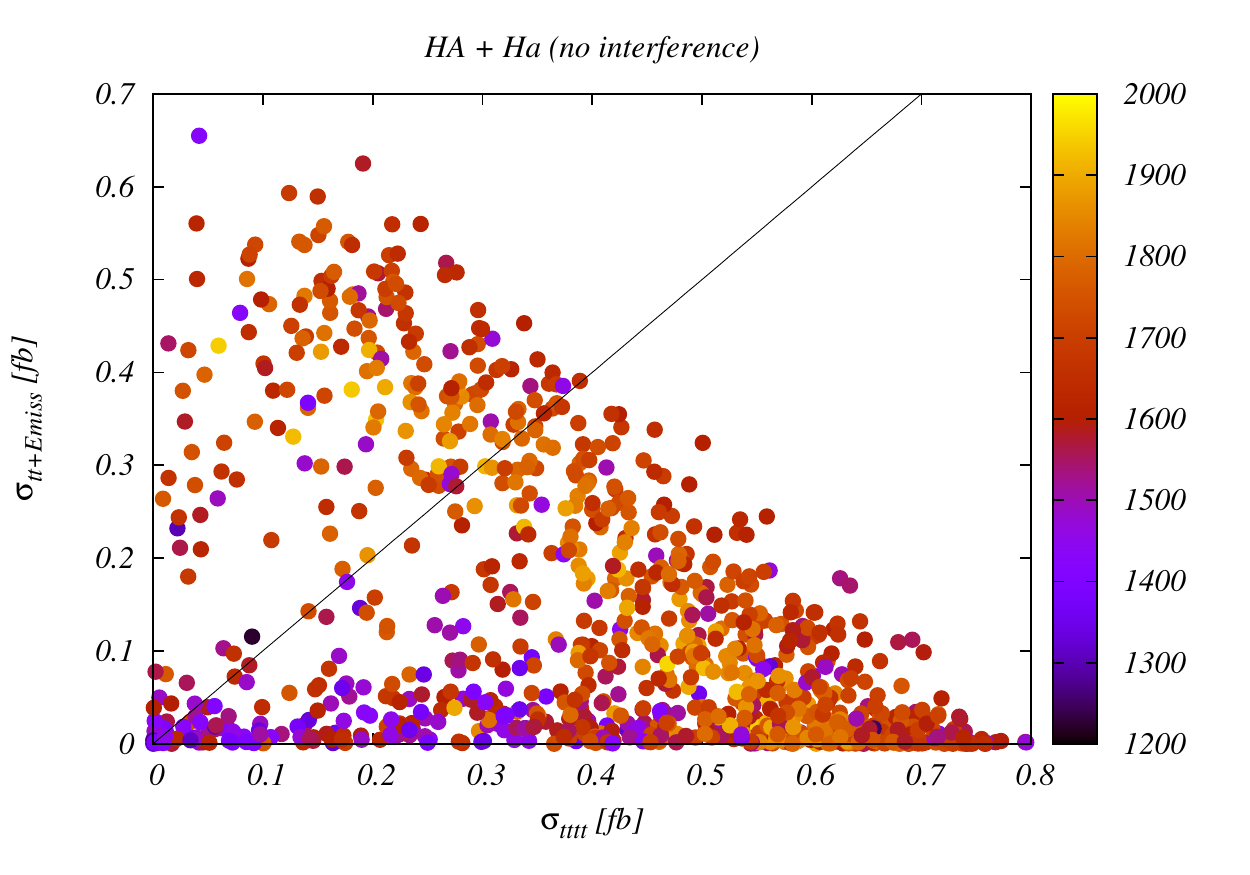}
\end{center}
\end{minipage}
\end{center}
\caption{\label{fig:thdmaatee} Production cross sections for $t\bar{t}t\bar{t}$ (x-axis) and $t\bar{t}+\slashed{E}$ (y-axis) final state in a factorized approach, for an $e^+e^-$ collider with a 3 \TeV center-of-mass energy. {\sl Left:} mediated via $HA$, {\sl right:} mediated via $HA$ and $Ha$ intermediate states. Color coding refers to $m_H+m_A$ {\sl (left)} and $M_H+0.5\times\,\lb m_A+m_a\rb$ {\sl(right)}. Figures taken from \cite{Robens:2021lov}, with an update including results from \cite{ATLAS-CONF-2021-029}.}
\end{figure}
\end{center}

I also present cross section values for production at $e^+e^-$ colliders for points that pass all bounds considered in \cite{Robens:2021lov} at a 3 \TeV~ collider in figure  \ref{fig:thdmaatee}.

\section{Conclusion}
I briefly presented two scenarios for models with dark matter candidates and their prospective signatures and rates/ discovery ranges at future lepton colliders, with a focus on larger $\lb \mathcal{O}\lb \TeV\rb\rb$ center-of-mass energies. For the IDM, a detailed study shows that many still viable parameter points should be accessible, depending on the specifics of the particular benchmark points. For the THDMa, regions in parameter space exist where $t\bar{t}+\slashed{E}$ is the dominant production mode.

\end{document}